\documentclass[twocolumn]{emulateapj}
\usepackage{graphicx}
\usepackage[dvips]{color}

\newcommand{\be}{\begin{equation}}
\newcommand{\ee}{\end{equation}}
\newcommand{\ba}{\begin{eqnarray}}
\newcommand{\ea}{\end{eqnarray}}

\def\ap{\approx}

\def\lsim{\raise0.3ex\hbox{$\;<$\kern-0.75em\raise-1.1ex\hbox{$\sim\;$}}}
\def\gsim{\raise0.3ex\hbox{$\;>$\kern-0.75em\raise-1.1ex\hbox{$\sim\;$}}}

\def\theta{\vartheta}

\def\F{{\cal F}}

\def\gr{$\gamma$-ray}
\def\ES{1ES 0229+200}
\shortauthors{Dolag et al.}
\shorttitle{Lower limit on extragalactic magnetic fields}

\begin{document}
\author{K.~Dolag$^{1}$, M.~Kachelriess$^{2}$, S.~Ostapchenko$^{2,3}$, 
R.~Tom\`as$^{4}$}
\affil{
$^{1}$Max-Planck-Institut f\"ur Astrophysik, Garching, Germany\\
$^{2}$Institutt for fysikk, NTNU, Trondheim, Norway\\
$^{3}$D. V. Skobeltsyn Institute of Nuclear Physics, Moscow State University, 
Russia\\
$^4$II.~Institut f\"ur theoretische Physik, Universit\"at Hamburg, Germany}

\title{Lower limit on the strength and filling factor of extragalactic magnetic fields}

\begin{abstract}
High energy photons from blazars can initiate electromagnetic pair cascades
interacting with the extragalactic photon background. The charged component of 
such cascades is deflected and delayed by extragalactic magnetic 
fields (EGMF), reducing thereby the observed point-like flux and leading 
potentially  to multi degree images in the GeV energy range. We calculate 
the fluence of \ES\  as seen
by Fermi-LAT for different EGMF profiles using a Monte Carlo simulation 
for the cascade development. The non-observation of \ES\ by 
Fermi-LAT suggests that the EGMF fills at least 60\% of space
with fields stronger than ${\cal O}(10^{-16}-10^{-15})$\,G for life times of
TeV activity of ${\cal O}(10^2-10^4)$\,yr.
Thus the (non-) observation of GeV extensions
around TeV blazars probes the EGMF in voids
and puts strong constraints on the origin of EGMFs: Either EGMFs 
were generated in a space filling manner (e.g.\ primordially) 
or EGMFs produced locally (e.g.\ by galaxies) 
have to be efficiently transported to fill a significant volume fraction, 
as e.g.\ by galactic outflows.

\end{abstract}


\maketitle

\section{Introduction}

While magnetic fields are known to play a prominent role for the 
dynamics and in the energy budget of astrophysical systems on 
galactic and smaller scales, their role on larger scales is still 
elusive~\citep{r02,r08}. 
Extragalactic magnetic fields (EGMF) are notoriously difficult to 
measure and the data are incomplete. So far only in a few galaxy 
clusters observational constraints have been obtained, either by
observing their synchrotron radiation halos or by performing Faraday 
rotation measurements (RMs). Within galaxy clusters the inferred magnetic 
fields are between 0.1 and 1.0\,$\mu$G on scales as large as 1 Mpc, 
and can be as strong as 30\,$\mu$G localized inside cluster cool cores. 
However, as both observational methods need a prerequisite to 
measure magnetic fields (high thermal density for RMs and presence of 
relativistic particles for radio emission), they have been successfully 
applied only to high density regions of collapsed objects as galaxies and 
galaxy clusters. Fields significantly below $\mu$G level are barely 
detectable with these methods. Also other constraints, for instance the 
absence of distortions  in the spectrum and the polarization properties 
of the cosmic microwave background  radiation implies only a fairly 
large, global upper limit on the EGMF at the level of  $10^{-9}$\,G.

The observed magnetic fields in galaxies and galaxy clusters are
assumed to result from the amplification of much weaker seed
fields. Such seeds could be created in the early universe, 
e.g.\ during phase transitions, and then amplified by plasma processes. 
Alternatively, an early population of starburst galaxies or AGN could 
have generated the seeds of the EGMFs at redshift between five
and six, before galaxy clusters formed as gravitationally
bound systems. 
In both cases, a large fraction of the material collapsing to today's
visible structures will be seeded by such fields, but whether a significant 
fraction of the volume of the universe can be filled by such EGMFs depends
on the efficiency of the transport processes involved.
A quite different possibility is that the ejecta of AGN magnetized the 
intracluster medium only at low redshifts, and that thus the EGMF are confined 
within galaxy clusters and groups. 
The detection of EGMFs outside clusters is therefore crucial in 
discriminating different models for the origin of their seed fields.

An alternative approach to obtain information about the EGMFs is 
to use its effect on the radiation from TeV gamma-ray sources.
The multi-TeV \gr\ flux from distant blazars is strongly attenuated 
by pair production on the infrared/optical extragalactic 
background light (EBL),  initiating electromagnetic cascades in the 
intergalactic space. The charged component of these cascades is 
deflected by the EGMF. Potentially observable effects of such 
electromagnetic cascades in the EGMF include the delayed ``echoes" of 
multi-TeV \gr\  flares or gamma-ray bursts~\citep{plaga,jap} and the 
appearance of extended 
emission around initially point-like \gr\ sources 
\citep{coppi,neronov07,halo1,el09,jet}.

An additional way to derive lower limits on the EGMF has been 
pointed out recently by \citet{ne10} and \citet{ta10}: Since the
deflection of the cascade flux into an extended halo weakens the
point-like image, the non-observation of TeV blazars in the GeV range
by Fermi-LAT can been used to derive a lower limit on the
EGMF. Particular suitable candidates are blazars with a very 
hard TeV spectrum like \ES\ that show a low
intrinsic GeV emission. In this way, \citet{ne10} and \citet{ta10}
derived the lower bound $B\gsim 5\times 10^{-15}$\,G on the EGMF.
Note that \citet{ak} found evidence for gamma-ray halos in the 
stacked images of the 170 brightest active galactic nuclei 
observed with Fermi-LAT. The size of these halos are consistent with  
$B\sim 10^{-15}$\,G, however it is unclear if the excess found is real
or due to the imperfect knowledge of the Fermi-LAT point spread function 
(PSF) as argued by \citet{Neronov:2010bi}.
All these analyses assumed a stationary source and did not discuss
the effect of time delays induced by EGMFs.

In this work, we will concentrate on \ES\ and follow
essentially the same assumptions about the source and the sensitivity
of Fermi-LAT as \citet{ta10}. We will improve on previous analyses in
two respects: First, we use a Monte Carlo simulation for the
development of electromagnetic cascades in the EBL that includes the
effects of magnetic fields like synchrotron radiation and deflections
of electrons. Moreover, we calculate the time delay that is
induced by the EGMF deflections.
Second, we examine the influence of a more realistic,
structured magnetic field on the EGMF limit.

\section{Simulation procedure}

We inject at the redshift $z=0.14$ photons with an energy
distribution given by $\F\propto E^{-2/3}$ and a maximal
energy $E_{\max}=20$\,TeV. Such a hard injection spectrum gives a 
reasonable fit to the HESS data and is consistent with recent 
SWIFT~\citep{swift} observations as argued by \citet{ta09}.
We assume also  a relatively low Lorentz factor, $\Gamma=10$, 
i.e.\  $\Theta_{\rm jet}=6^\circ$, in order to derive a conservative
lower bound on the EGMF.

We follow electromagnetic cascades initiated by these high energy photons with
the Monte Carlo code introduced by \citet{CenA}. Since Fermi-LAT provides 
only an upper limit on the photon flux from \ES\, it is sufficient to 
assume a radial-symmetric halo and to calculate the deflections
and time delays
as described in~\citep{halo1}. We use as EBL background the best-fit
model from the calculations of \citet{kneiske10}.

We divide the photons arriving at the observer into a point-like
flux and a halo outside a circle with angle $\theta_{\ast}$ around the 
source. For Fermi-LAT, we use for $\theta_{\ast}$ an analytical 
approximation of the angle $\theta_{95}$ containing 95\% of the
flux emitted by a point-source
given by
%
$ \theta_{95}\ap 1.68^\circ (E/{\rm GeV})^{-0.77} + 0.2^\circ
 \exp(-10\,{\rm GeV}/E)$.  
%
Above 300 GeV, $\theta_{95}$ is set equal to
$0.11^\circ$, a value typical for the resolution of atmospheric 
Cherenkov telescopes.

\section{Results for different EGMF models}

\begin{figure}
\vskip-2.2cm
\includegraphics[width=\linewidth,angle=0]{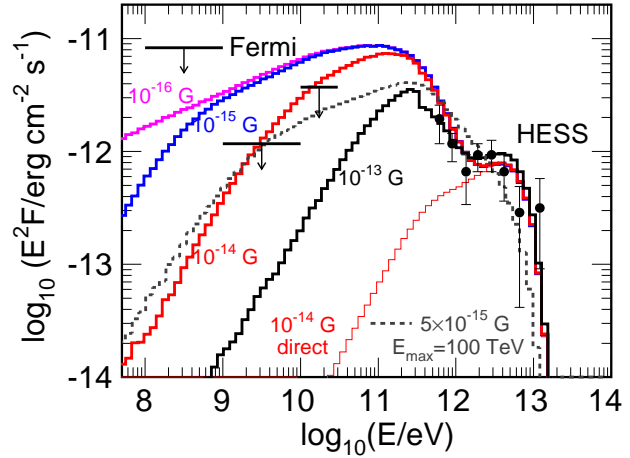}
\caption{Fluence contained inside the 95\% confidence contour of the 
PSF of  Fermi-LAT as function of energy together with Fermi-LAT upper 
limits and HESS observations for a uniform EGMF with strengths
varying from $B=10^{-16}$\,G  to $B=10^{-13}$\,G with
 $E_{\rm max}=20$\,TeV (solid) and 100 TeV (dashed). The direct
  component for $B=10^{-14}$\,G is also shown.}
\label{fig:uni}
\end{figure}

\begin{figure}
\vskip-2.2cm
\includegraphics[width=\linewidth,angle=0]{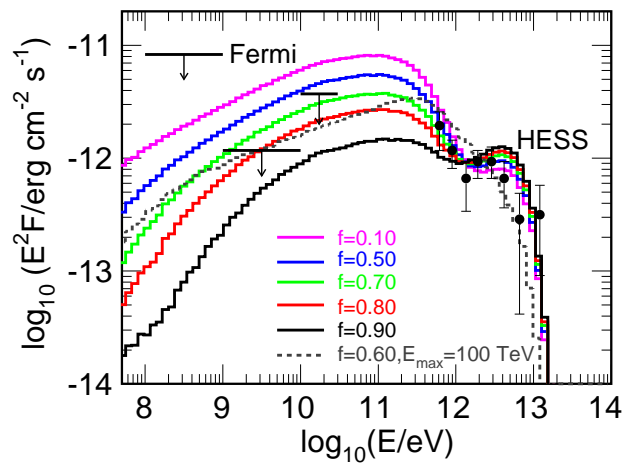}
\caption{
Fluence contained inside the 95\% confidence contour of the 
PSF of  Fermi-LAT as function of energy for a EGMF with top-hat
profile and filling factor $f$ varying from $f=0.1$ to $f=0.9$ 
with $E_{\rm max}=20$\,TeV (solid) and 100 TeV (dashed).} 
\label{fig:fill}
\end{figure}

In Figs.~\ref{fig:uni} and \ref{fig:fill} we show our results for the
fluence contained inside the 95\% confidence contour of the PSF of 
Fermi-LAT. Additionally, these figures contain the HESS observations
\citep{HESS} as black dots with error bars and the Fermi-LAT upper 
limits derived by \citet{ta10}. The fluences have been
normalized fitting them to the HESS data.

In Fig.~\ref{fig:uni}, we have used a uniform magnetic field so that
our results can be directly compared with the analytical estimates of
\citet{ta10}.  Note that a turbulent field with correlation length
$L_{\rm cr}$ much larger than the mean free path $l_{\rm IC}$ of
electrons in the Thomson regime, $L_{\rm cr}\gg l_{\rm IC}\sim
1$\,kpc, is well approximated by a
uniform field. For smaller correlation lengths, $L_{\rm cr}\ll l_{\rm
  IC}$, the electron diffuses in the small-angle deflection regime,
requiring larger magnetic fields for the same deflection angle.
Demanding that the cascade flux is below the upper limits of Fermi-LAT
leads to a lower limit on the magnetic field strength of
$\sim 10^{-14}$\,G. For this case the direct component, i.e.\ 
photons arriving at the detector without cascading, is also
  shown.
Note that for small $E_{\max}$ the transition from the direct to the 
cascade contribution leads  to a break at $\sim$\,TeV in the 
spectrum, as suggested by the HESS data.

While our limit agrees
reasonably well with the analytical estimate of \citet{ta09}, the
shape of the cascade flux obtained differs. There are several reasons
responsible for these differences: First, \citet{ta09} assume that the
spectral shape of the cascade flux below the threshold energy $\sim
10^{11}$\,eV is given by $\F\propto E^{-0.5}$ for negligible magnetic
fields. Such a slope typical for the regime of Thomson cooling is
however restricted to energies $E\lsim 10^8$\,eV, while at higher
energies a plateau $\F\propto E^\alpha$ with $\alpha\sim -0.9$ is
expected~\citep{casc}. Second, deflections in the EGMF lead even for
isotropic emission to extended images of point-like sources.  This
effect has been neglected in \citet{ta09}.  Third, using full
probability distributions for the interactions there is a
non-negligible probability for photons not to
interact, especially towards the low-energy end of the
  injected energy range.  Finally, the energy dependent PSF of Fermi
introduces an artificial energy dependence of the point-like flux
$\F\propto \theta_{95}^2$.
Note that an increase in $E_{\rm max}$ from 20 to 100 TeV {\em reduces\/} 
the limit on the magnetic field strength to $\sim 5\times 10^{-15}$\,G, 
see Fig.~1, while a further increase of $E_{\rm max}$ strengthens the limit
again. The counter-intuitive behavior between 20 to 100\,TeV is caused
by the dominance of direct photons at the high-energy part of the
spectrum for small $E_{\max}$.

We discuss next the consequence of time delays induced by the
  EGMF.  We stress that the non-observation of \ES\ by Fermi-LAT is a
  signature for non-zero EGMFs, even if the delay of the Fermi signal
  would exceed the life time $\tau$ of \ES\ as TeV source. However,
  the numerical value of the limit deduced using deflections would be
  modified. For $B=5\times 10^{-15}$\,G, we find as average time delay
  $1\times 10^6$, $1\times 10^5$, and 5000\,yr for the three energy
  bins (0.1--10), (1--10), and (10--30)\,GeV. Thus the high-energy bin
  of the Fermi data can be used to derive the constraint as above, if
  \ES\ is a relatively stable source over a time scale longer than
  $10^4$\,yr.  Even if $\tau$ is much smaller, say around 100\,yr, the
  lower limit on $B$ weakens only by a factor 10 to $B={\cal
    O}(10^{-16}\,{\rm G})$.

Since the EGMF is strongly structured, one may wonder how  a non-uniform  
field modifies this limit. In particular, we want to
address the question whether the presence of relatively strong fields
concentrated inside cosmic structures like filaments could mimic the
effect of an EGMF present also in voids. As simplest possible test,
we use first a top-hat profile for the structure of the EGMF: 
We set the field strength to zero in a fraction $1-f$ of space and use a value 
which in general is assumed to be representative for filaments,
$B=10^{-10}$\,G, in the remaining part. For the
separation of the peaks we use $D=10$\,Mpc motivated by the typical 
distances between cosmological structures, 
although the exact value of $D$ plays
no role as long as $(1-f)D\ll l_\gamma$, with  $l_\gamma$ as
mean free path of photons. 
The dependence of the fluence contained inside the PSF of Fermi-LAT
on the filling factor $f$ is shown in Fig.~\ref{fig:fill}. To be 
consistent with the Fermi upper limits, sufficiently strong magnetic 
fields should fill $\gsim 80\%$ of space. The derived limit on the
filling factor is practically independent of the source life time $\tau$ and 
$B$, as long as the
field is stronger than $\gtrsim 5\times 10^{-15}$\,G.
As in the previous case, by assuming a higher injected $E_{\rm max}$ the
required filling factor is slightly reduced to 60\% . 

The failure of strong fields filling only a small fraction of the universe
to suppress sufficiently the point-like cascade flux can be understood
as follows: The HESS observations of \ES\ cover the energy range 
0.5--11\,TeV. In the same energy range, the mean free path $l_\gamma$ 
of VHE \gr s through the EBL varies between 1000 and 50\,Mpc
and is thus always much larger than the typical extension of
regions with large fields, $(1-f)D$. For the energies considered,
the cascade consists typically of only three steps, 
$\gamma\to e^\pm\to\gamma$. Since the mean free path $l_{\rm IC}$ 
of electrons in the Thomson regime is very small, $l_{\rm IC}\sim 1$\,kpc, 
all cascades with electrons created outside the strong-field regions are
undeflected. Thus it is not possible to trade smaller values of
$f$ against larger values of $B$: Increasing the field strength
beyond $\sim 10^{-13}$\,G leads only to an increase of the deflection,
while the fraction of cascades deflected outside the Fermi PSF
remains constant.

\begin{figure}
\vskip-2.2cm
\includegraphics[width=\linewidth,angle=0]{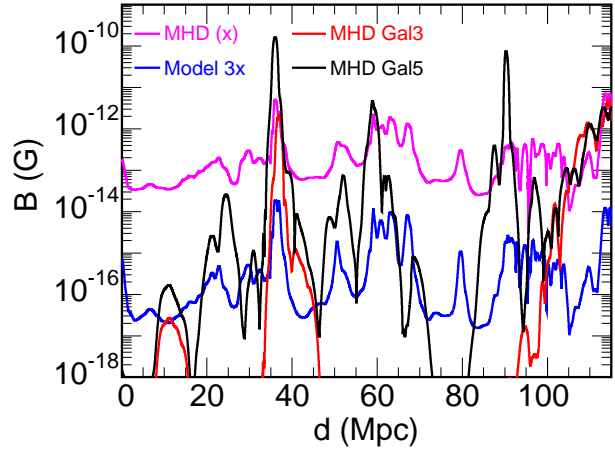}
\caption{The EGMF component perpendicular to the line-of-sight
towards \ES\ from four different MHD simulations.}
\label{fig:Bprofiles}
\end{figure}

To mimic the even more complex situation in presence of cosmological 
structures, we have used magnetic fields derived from several 
cosmological MHD simulations using constrained initial conditions
(Dolag et al. 2004, Donnert et al. 2009; for alternative simulations
see e.g.\ Ryu et al. 2008). Therefore they reproduce fairly 
well the true large scale matter distribution, out to a distance of 
$\approx 114$\,Mpc, cf.\  for more details Dolag et al. (2005).
We extracted the EGMF
as predicted by these simulations along the line of sight towards the 
position of 1ES 0229+200 for several models for the magnetic seed field.
A comparison
of the EGMF component perpendicular to the line-of-sight
towards \ES\ obtained in the models is shown
in Fig.~\ref{fig:Bprofiles}.

\begin{figure}
\vskip-2.2cm
\includegraphics[width=0.9\linewidth,angle=0]{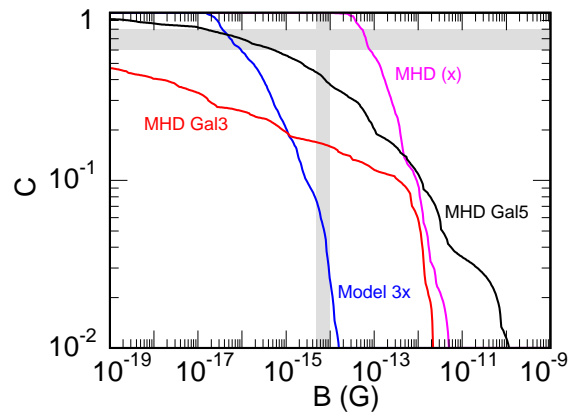}
\caption{Cumulative volume filling factor $C(B)$ 
for the four different EGMF models found in MHD simulations.
\label{fig:Bcumulative}}
\end{figure}

\begin{figure}
\vskip-2.2cm
\includegraphics[width=\linewidth,angle=0]{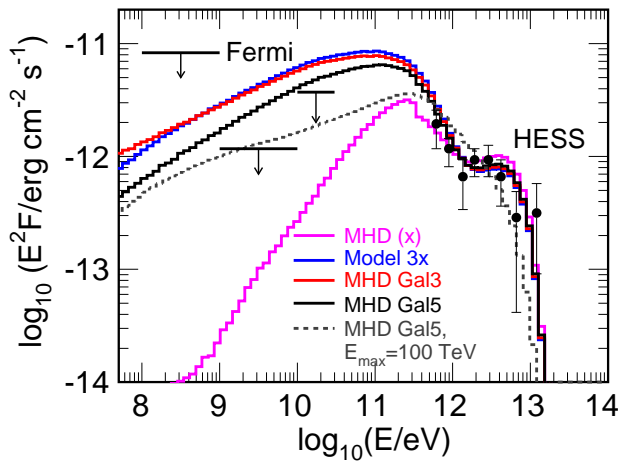}
\caption{Fluence contained inside the 95\% confidence contour of the 
PSF of  Fermi-LAT as function of energy for EGMFs from 
four different MHD simulations  with
  $E_{\rm max}=20$\,TeV (solid) and 100 TeV (dashed).}
\label{fig:Bresult}
\end{figure}

Most magnetic seed models used in cosmological simulations 
assumed a space filling (e.g.\ primordial) magnetic seed field of order 
$10^{-12}$G, which inside galaxy clusters is amplified by the collapse 
and additionally by shear turbulence up to the observed $\mu$G level. For the 
purpose of this work 
we have selected the available simulation with the lowest primordial seed 
field of $10^{-13}$\,G (MHDx), which inside voids is even diluted to values 
of order of $10^{-14}$\,G due to cosmic expansion. To construct a second model 
with magnetic fields as low as $10^{-17}$\,G inside voids, 
we scaled down the magnetic field of the original simulation by a factor 
proportional to the local density to the power 2/3 (Model 3x) with the 
fix point at the core density of galaxy clusters. Thereby, the Faraday 
RMs of the newly constructed model within galaxy 
clusters does not significantly change, whereas the magnetic field in 
filaments and voids drop by nearly three orders of magnitude compared 
to the original simulation.

Models which intrinsically produce lower magnetic fields (and filling
factors) in filaments and voids are those where the seed
field originates from the galaxies formed. In
such simulations, galaxies are identified at early times, and their
halos are magnetized according to the magnetic field observed in
nearby galaxies.  So far such simulations do not include any transport
processes for the magnetic field (like galactic outflows or AGN
activity) beyond the large scale potential well in which galaxies
live. 
Nevertheless, these models reproduce well the
  observed magnetic field properties within galaxy clusters, since
  the gas component of the galactic halos is stripped
  during the formation process of the cluster and the magnetic field
  gets well mixed and subsequently amplified by the turbulence within
  the intercluster medium.
The result of such a simulation is the
model (MHD Gal3) from Donnert et al.\ 2009. As
galaxies in filaments form later than their counterparts in high
density environment, we have also extracted the magnetic field from a
model which subsequently put magnetic fields into newly formed halos
also at lower redshift (MHD Gal5). The main difference to the previous
model is the presence of more magnetized structures within filaments
whereas the magnetic field properties in galaxy clusters do not change
significantly.

The simulation used covers only up to distances of
114\,Mpc, and therefore we are forced to extend the magnetic field
profile beyond that point. For this purpose we assume
that the extended field has the same statistical properties
as the simulated patch.
In practice, we simply mirror and repeat the field beyond the
first 114\,Mpc.
In Fig.~\ref{fig:Bcumulative}  we show
  the cumulative volume filling factor, i.e.\ the fraction $C(B)$ of
  volume occupied by regions with field strength $>B$. The vertical
  and horizontal bands stand for the minimum magnetic field and
  filling factor, respectively, required to avoid the Fermi limit, as
  obtained in Figs.\,\ref{fig:uni} and \ref{fig:fill}.

The fluence contained inside the PSF of Fermi-LAT for the EGMF models
MHDx, MHD Gal3 and MHD Gal5 is shown in Fig.~\ref{fig:Bresult}.  From
the results for the top-hat model, we anticipate that the EGMF in the
models MHD3x and MHD Gal3 is too weak and does not lead to a
significant modification of the fluence, see
  Fig.~\ref{fig:Bcumulative}: 
Since the model
  MHD Gal5 is not far from the intersection of the two bands, 
  we expect it not to lie far from the Fermi limit. In
  Fig.~\ref{fig:Bresult} it is seen how in the particular case of
  $E_{\rm max}=100$\,TeV this model is marginally consistent with the Fermi-LAT
  observations.

Finally, the fluence is several orders of magnitude below the
observational limits for fields as strong as in the model MHDx. 
In such a case, extended images in the GeV range as predicted e.g.\
by \citet{jet} will be impossible to observe with Fermi-LAT.

We would like to point out, that on the other hand, all these models 
are in agreement with observations of magnetic fields in galaxy clusters -- 
which 
so far is the only place where the strength and structure of magnetic fields 
has been directly detected -- and therefore demonstrate the constraining 
power of using high energy photons to infer the origin and the processes 
leading tho EGMFs.

\section{Summary}

We have calculated the fluence of \ES\ as seen by Fermi-LAT using a
Monte Carlo simulation for the cascade development.  We have discussed
the effect of different EGMF profiles on the resulting suppression of
the point-like flux seen by Fermi-LAT.  Since the electron cooling
length is much smaller than the mean free path of the TeV photons, a
sufficient suppression of the point-like flux requires that the EGMF
fills a large fraction along the line-of-sight towards \ES,
$f\gsim 0.6$. The lower limit on the magnetic field
strength in this volume is $B\sim {\cal O}(10^{-15})$\,G,
assuming \ES\ is stable at least for $10^4$\,yr, weakening by a factor of
10 for $\tau=10^2$\,yr. 
These limits put very stringent constraints on the
origin of EGMFs: Either the seeds for EGMFs have to be produced by a
volume filling process (e.g.\ primordial) or very efficient transport
processes have to be present which redistribute magnetic fields that
were generated locally (e.g.\ in galaxies) into filaments and voids
with a significant volume filling factor.

\vskip0.2cm \textit{Acknowledgments.}  We are grateful to L. Costamante and
D. Paneque for discussions and to the referee
for his/her useful remarks. K.D.\ is supported partly by
the DFG SFB 1177 and a Cluster of Excellence, while S.O.\ acknowledges
a fellowship from the program Romforskning of Norsk
Forsknigsradet. 

\vskip0.2cm
\textit{Note added.} 
After the completion of this work the preprint of Tavecchio et al.\ (2010b)
appeared in which these authors study the four BL Lac objects (RGB J0152+017, 
1ES 0229+200, 1ES 0347-121 and PKS 0548-322) and derive a bound of
$B\sim 5\times 10^{-15}$\,G on a space filling magnetic field.


\end{document}